\documentclass[pra,twocolumn,showpacs,floatfix,superscriptaddress]{revtex4}

\usepackage{amsmath,amssymb,amsthm}
\usepackage{times}
\usepackage{graphicx}
\usepackage{bm}
\usepackage[dvips]{color}

\begin{document}

\title{Quantum chaos and thermalization in gapped systems}

\author{Marcos Rigol} 
\affiliation{Department of Physics, Georgetown University, Washington, DC 20057, USA} 
\author{Lea F. Santos}
\affiliation{Department of Physics, Yeshiva University, New York, New York 10016, USA}


\begin{abstract}
We investigate the onset of thermalization and quantum chaos in finite one-dimensional gapped 
systems of hard-core bosons. Integrability in these systems is broken by next-nearest-neighbor 
repulsive interactions, which also generate a superfluid to insulator transition. 
By employing full exact diagonalization, we study chaos indicators and few-body observables.
We show that with increasing system size, chaotic behavior is seen over a broader range of 
parameters and, in particular, deeper into the insulating phase. Concomitantly, we observe that, 
as the system size increases, the eigenstate thermalization hypothesis extends its range 
of validity inside the insulating phase and is accompanied by the thermalization of the system.
\end{abstract}

\maketitle


The use of ultracold quantum gases to realize strongly correlated phases of matter has been 
at the forefront of research during the last decade \cite{bloch08}. Such systems will not 
only help to understand phases and models introduced in the past, but also will create and 
will provide the means to investigate new exotic phases. Another major promise of the field is 
the possibility for studying the dynamics of correlated quantum systems in very controlled 
ways. For short times, the collapse and revival of phase coherence was analyzed in 
Ref.\ \cite{greiner02b} after an interaction quench from the Bose-Einstein condensate to 
the Mott insulating regime. {\it Ab initio} calculations of various bosonic and fermionic 
lattice models reproduced those observations \cite{rigol06STATb,kollath07,manmana07}. A 
natural question that follows, given the fact that these systems are nearly isolated, 
is whether conventional statistical ensembles can describe experimental observables after 
relaxation. 

Recent experiments have found that relaxation toward thermal equilibrium takes place in certain 
setups, but not in others \cite{kinoshita06}. The lack of thermalization in the latter case 
can be associated with the proximity to integrability \cite{rigol09STATa}. At integrability, 
several works have shown that thermalization is not expected to occur \cite{rigol07STATa}, 
except for special conditions and/or observables \cite{rigol06STATb,rossini09}. For generic 
nonintegrable systems, thermalization is expected to occur and follows from the eigenstate 
thermalization hypothesis (ETH) \cite{deutsch91,rigol08STATc}.
 
Interestingly, in numerical calculations for quenches across a gapless to gapped phase, 
thermalization did not occur in the gapped side, even when the system was nonintegrable
\cite{kollath07}. As a matter of fact, ETH, which was shown to hold in several nonintegrable 
gapless systems \cite{rigol08STATc,rigol09STATa}, has been questioned for gapped systems 
\cite{biroli09,roux09a}. Questions raised include the proximity to the atomic limit, 
finite-size effects \cite{roux09a}, and the effects of rare states \cite{biroli09} within the 
insulating phase. Other studies that deal with spin and fermionic systems have found that 
relaxation toward equilibrium occurs faster close to a critical point \cite{eckstein09}. The 
fact that, away from the ground state, these systems are, in general, not insulating, even 
if gaps are present in the spectrum, renders the debate more interesting still. Why then 
would generic nonintegrable gapped systems behave differently from gapless ones? Remarkably, 
in disordered systems, insulating behavior may take place away from the ground state \cite{huse}, 
and those ones could indeed behave differently.

In this Rapid Communication, we use various measures, such as quantum chaos indicators \cite{santos10} 
and eigenstate expectation values of experimental observables \cite{rigol09STATa} to 
understand whether thermalization should occur in gapped systems. This question is of interest 
for current experiments with ultracold gases, where systems with insulating ground states such 
as the bosonic and fermionic Mott insulators are studied. We show that thermalization does occur 
in the gapped side of the phase diagram and that, as the system size increases, thermalization 
is observed deeper into that phase. We also find that ETH holds for systems with larger gaps, 
as the system size is increased. This supports the view that thermalization occurs in generic 
nonintegrable systems independent of the presence or absence of gaps in the spectrum. One does 
need to be careful with temperature effects and finite-size effects, which may be more relevant 
close to integrable points \cite{rigol09STATa} and in systems with gaps \cite{roux09a}. We also 
study the long-time dynamics of these gapped systems and address its universality.

We focus our study on the one-dimensional hard-core boson (HCB) model with nearest-neighbor 
hopping $t$, and nearest- and next-nearest-neighbor (NNN) interaction $V$ and $V'$, respectively. 
The Hamiltonian is written as
{\setlength\arraycolsep{0.5pt}
\begin{eqnarray}
&&\hat{H}_{b}=\sum_{i=1}^L \left\lbrace -t\left( \hat{b}^\dagger_i \hat{b}_{i+1} + 
              \textrm{H.c.} \right) \right. \label{Eq:hamiltonianHCB} \\
&&+V\left.\left( \hat{n}^b_i-\dfrac{1}{2}\right)\left( \hat{n}^b_{i+1}-\dfrac{1}{2}\right) 
 +V'\left( \hat{n}^b_i-\dfrac{1}{2}\right)\left( \hat{n}^b_{i+2}-
  \dfrac{1}{2}\right)\right\rbrace \nonumber
\end{eqnarray}
}where standard notation has been used \cite{rigol09STATa}. We restrict our analysis to lattices 
with 1/3 filling ($N_b=L/3$); $t=1$ sets the energy scale, $V=6$, and $V'$ is varied 
($0\leq V'\leq9$). For $V'<V'_c=3$, the ground state is a gapless superfluid, whereas for 
$V'\geq V'_c=3$, it is a gapped insulator \cite{zhuravlev97}.

Eigenstate expectation values (EEVs) of different observables, as well as the nonequilibrium 
dynamics and thermodynamics of these systems, are determined using full exact diagonalization 
of the Hamiltonian in Eq.\ (\ref{Eq:hamiltonianHCB}). We study lattices with up to 24 sites and 8 HCBs, 
which correspond to a total Hilbert space of dimension $D=735\,471$. We take advantage of the 
translational symmetry of the lattice to independently diagonalize each Hamiltonian block with 
total momentum $k$; the largest block in $k$ space has dimension $D_k=30\,667$ \cite{supplement}.

The Hamiltonian in Eq.\ (\ref{Eq:hamiltonianHCB}) is integrable when $V'=0$, whereas the addition of NNN 
interaction leads to the onset of chaos. Note that the integrable-chaos transition may occur 
although random elements are nonexistent in the Hamiltonian \cite{santos10}. We start our 
study by addressing how quantum chaos indicators, such as level spacing distribution, level 
number variance, and inverse participation ratio (IPR) \cite{supplement}, change as one moves 
away from the integrable point and eventually enter the gapped region by increasing $V'$. 
The outcomes are shown to support our results on thermalization.

Level spacing distribution and level number variance are obtained from the unfolded spectrum
of each $k$ sector separately. The first is a measure of short-range correlations, and the 
second is a measure of long-range correlations~\cite{Guhr1998}. For integrable systems, the distribution of 
spacings $s$ of neighboring energy levels may cross, and the distribution is Poissonian
$P_{P}(s) = \exp(-s)$, while for nonintegrable systems, level repulsion leads to the 
Wigner-Dyson distribution. The form of the latter depends on the symmetries of the system. 
Here, it coincides with that of ensembles of random matrices with time reversal invariance, 
the so-called Gaussian orthogonal ensembles (GOEs): $P_{\text{GOE}}(s) = (\pi s/2)\exp(-\pi s^2/4)$.
The level number variance is defined as
$\Sigma^2(l) \equiv \langle N(l,\epsilon)^2 \rangle - \langle N(l,\epsilon)\rangle^2$,
where $N(l,\epsilon)$ is the number of states in the energy interval $[\epsilon,\epsilon+l]$
and $\langle . \rangle $ is the average over different initial values of $\epsilon$.
For a Poisson distribution $\Sigma^2_{P}(l)=l$, while for GOEs in the limit of large $l$,
$\Sigma^2_{\text{GOE}}(l)=2[\ln (2\pi l) + \gamma +1 -\pi^2/8 ]/\pi^2$, where $\gamma$ is the 
Euler constant.

\begin{figure}[htb]
\includegraphics[width=0.47\textwidth]{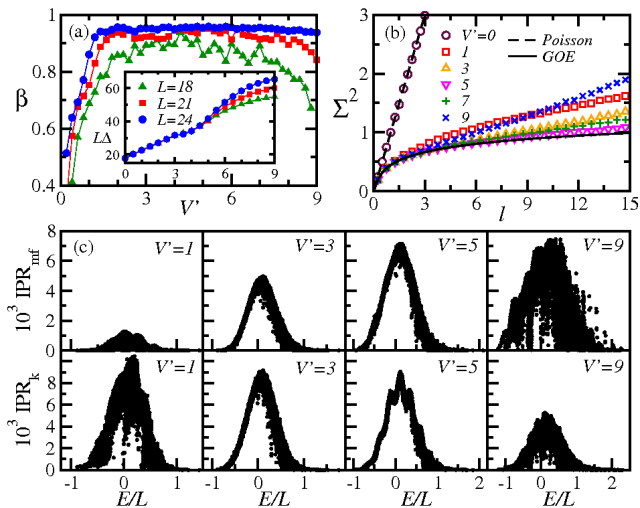}
\vspace{-0.25cm}
\caption{(Color online)
(a) Parameter $\beta$ of the Brody distribution used to fit the average of the level 
spacing distributions over all sectors with $k=1, \ldots, \left\lfloor L/2 \right\rfloor$. 
The inset shows the gap (between the lowest energy states) times $L$ vs $V'$. (b) Level 
number variance averaged over the same $k$ sectors, for $L=24$, and compared to the 
Poisson and GOE results. (c) Inverse participation ratio in the mean-field (mf) (top) and 
momentum (bottom) basis vs energy per site; $L=24$, $k=2$, and $D_k=30\,624$.}
\label{fig:chaos}
\end{figure}

Figures \ref{fig:chaos}(a) and  \ref{fig:chaos}(b) show results for $P(s)$ and $\Sigma^2(l)$, 
respectively, for different values of $V'$. The level spacing distribution is parametrized 
by $\beta$, which is used to fit $P(s)$ with the Brody distribution~\cite{Brody1981},
$P_B(s) = (\beta +1) b s^{\beta} \exp \left( -b s^{\beta +1} \right)$, where
$b= \left[\Gamma \left( \frac{\beta + 2}{\beta +1} \right)\right]^{\beta +1}$ \cite{supplement}.
Two transitions are verified: (i) from integrable [$\beta\rightarrow 0, \Sigma^2(l)  \rightarrow l$] 
to chaotic [$\beta\rightarrow 1, \Sigma^2(l)  \rightarrow \Sigma^2_{\text{GOE}}(l)$] 
as $V'$ increases from $V'=0$, and (ii) a departure from chaoticity for large values of $V'$. 
Figure~\ref{fig:chaos}(a) also shows a strong dependence of the results on the system size. 
For larger systems, smaller values of $V'$ suffice for the first integrable-chaos transition, 
and larger values of $V'$ are required for the second transition. It is an open question whether, 
in the thermodynamic limit, any value of $V'\neq0$ would implicate chaoticity for these systems. 
However, for our finite systems, we do find an overlap between the gapped phase and the chaotic 
regime. The behavior of the gap $\Delta$ (times $L$) vs $V'$ is depicted in the inset in 
Fig.~\ref{fig:chaos}(a). Notice the kink around $V'_c\sim3$, which signals the onset of the 
superfluid-insulator transition.

The IPR measures the level of delocalization of the eigenstates \cite{Izrailev1990}. Contrary 
to the two previous quantities, IPR depends on the basis chosen for the analysis. For an eigenstate 
$|\psi_{\alpha}\rangle$ of Eq.\ (\ref{Eq:hamiltonianHCB}) written in the basis vectors $|\phi_j\rangle$ 
as $|\psi_{\alpha}\rangle = \sum_{j=1}^{D_{k}} c^j_{\alpha} |\phi_j\rangle$, we have 
$\mbox{IPR}_{\alpha} \equiv (\sum_{j=1}^{D_k} |c^j_{\alpha}|^4)^{-1}$. In Fig.~\ref{fig:chaos}(c), 
we investigate IPR in two bases: the mf basis (IPR$_{\text{mf}}$), where the 
$|\phi_j\rangle$'s are the eigenstates of the integrable Hamiltonian ($V'=0$); and the $k$ basis 
(IPR$_{k}$), where the $|\phi_j\rangle$'s are the total momentum basis vectors. Small vs large 
values of IPR$_{\text{mf}}$ separate regular from chaotic behavior \cite{ZelevinskyRep1996}
and signal delocalization during the first integrable-chaos transition. The reduction of IPR$_{k}$ 
indicates localization in $k$ space, which in our model, results from approaching the atomic limit 
in the presence of translational symmetry. Hence, it explains the departure from 
chaoticity for $V'\gg V'_c$ \cite{supplement}. 

The structures of the eigenstates are intimately connected to the thermalization process. 
As stated by the ETH, thermalization is expected to occur when the expectation values of 
experimental observables (with respect to eigenstates of the Hamiltonian that are close in energy) 
are very similar to each other and, hence, are equal to the microcanonical average, that is, 
thermalization occurs at the level of eigenstates. This is certainly the case with GOEs, where 
the amplitudes $c^j_{\alpha}$ for all $|\psi_{\alpha}\rangle$'s are independent random numbers. 
GOE eigenstates in any basis then lead to $\mbox{IPR}_{\text{GOE}} \sim D_k/3$~\cite{ZelevinskyRep1996}. 
For our lattices, IPR$_{\text{mf}}$ and IPR$_{k}$ may approach the GOE value only in the middle 
of the spectrum, as expected for systems with finite-range interactions~\cite{Kota2001,santos10}.
States at the edges, even in the chaotic limit, are more localized. In addition, as shown 
in the panels of Fig.~\ref{fig:chaos}(c), the values of IPRs for eigenstates close in energy 
fluctuate significantly as one moves away from the chaotic limit, namely, as $V'\rightarrow 0$, 
where the system becomes localized in the mf basis, and for $V' \gg V'_c$, where the system 
becomes localized in the $k$ basis. Thus, ETH is not expected to hold in these regions. 
On the contrary, for intermediate values of $V'$, IPR$_{\text{mf}}$ and IPR$_{k}$ become smooth 
functions of energy, especially for $E<0$. We may, therefore, anticipate compliance with the ETH 
even after the opening of the gap. 

To verify the relation between ETH and the chaos measures analyzed previously, we have studied the 
EEVs of one- and two-body observables for different values of $V'$ as one crosses the superfluid 
to insulator transition. We have found qualitatively similar results for them, so we only report 
here on the kinetic energy [$K=\sum_i -t\left( \hat{b}^\dagger_i \hat{b}_{i+1} + \textrm{H.c.} \right)$] 
and the momentum distribution function [$n(k)$]. Both are one-body observables, although the former 
one is local, while the latter is not. $K$ and $n(k)$ are routinely measured in cold gases experiments.

\begin{figure}[htb]
\includegraphics[width=0.47\textwidth]{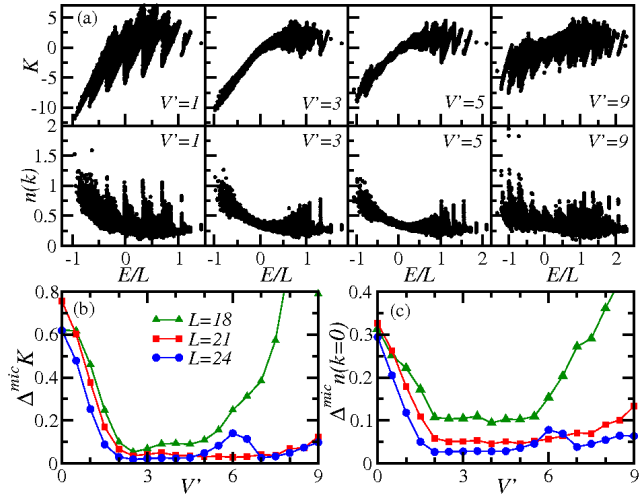}
\vspace{-0.25cm}
\caption{(Color online)
(a) EEVs of $K$ (top) and $n(k)$ (bottom) vs energy per site for 
the full spectrum (which include all momentum sectors). Results are shown for four different 
values of $V'$ and $L=24$. Panels (b) and (c) average relative deviation of the EEVs of (b) $K$ 
and (c) $n(k=0)$ with respect to the microcanonical result vs $V'$, for $T=3$ (see text) 
and three different lattice sizes. $T=3$ ($L=24$) corresponds to $E=-10.7$ for $V'=1$, 
$E=-10.0$ for $V'=3$, $E=-11.7$ for $V'=5$, and $E=-19.3$ for $V'=9$.}
\label{fig:eth}
\end{figure}

Figure \ref{fig:eth}(a) depicts the EEVs of $K$ and $n(k)$ for all eigenstates of the 
Hamiltonian and for different values of $V'$. For small values of $V'$ ($V'<2$ for $L=24$), 
there are large fluctuations of the EEVs of both observables over the entire spectrum. 
As $V'$ increases and one departs from integrability, these fluctuations reduce in the 
center of the spectrum, and ETH becomes valid. To increase $V'$ even further (beyond the 
superfluid to insulator transition), increases the fluctuations of the EEVs once again as 
the eigenstates begin to localize in $k$ space. These results are in clear agreement 
with what we expected based on the chaos indicators \cite{supplement}. 

To be more quantitative, we study the average deviation of the EEVs with respect to the 
microcanonical result ($\Delta^{mic}$). For an observable $O$, we define $\Delta^{mic}O=(\sum_{\alpha}\,
|O_{\alpha \alpha}-O_{mic}|)/(\sum_{\alpha}\,O_{\alpha \alpha})$, where the sum is performed over
the microcanonical window and $O_{\alpha\alpha}$ are the EEVs of $\hat{O}$. The microcanonical 
expectation values are computed as usual. We average over all eigenstates (from all momentum sectors) 
that lie within a window $[E-\Delta E, E+\Delta E]$, and take $\Delta E=0.1$. We have checked that 
our results are independent of the exact value of $\Delta E$ in the neighborhood of $\Delta E=0.1$. 
Here, we select $E$ such that, for different values of $V'$, the effective temperature ($T=3$) 
is the same for all systems sizes \cite{temperature}. 

Results for $\Delta^{mic}K$ and $\Delta^{mic}n(k=0)$ are presented in Figs.\ \ref{fig:eth}(b)
and \ref{fig:eth}(c) \cite{supplement}. They show that, in general, as the system size increases: 
(i) The average deviations of both observables decrease, and (ii) the upturn that occurs as 
localization starts to set in $k$ space moves toward larger values of $V'$. A comparison between
the lower and upper panels in Fig.\ \ref{fig:eth} also shows that where $\Delta^{mic}K$ and 
$\Delta^{mic}n(k=0)$ are minimal, so are the maximal fluctuations of $K$ and $n(k=0)$ in the 
individual eigenstates, and they decrease with increasing systems size \cite{supplement}. 
Hence, ETH is valid in that regime and we find no evidence of rare states 
\cite{biroli09}. The above results are in agreement with the chaos measure predictions 
and indicate that, for thermodynamic systems, ETH may be valid, away from the edges of the 
spectrum, even if one is deep into the insulating side of the phase diagram.

Equipped with this knowledge, we are now ready to study the dynamics of such systems after 
a quench. Our initial states are always selected from the eigenstates of (\ref{Eq:hamiltonianHCB}) 
with $t=1$, $V=6$, $V'_{ini}$ and zero total momentum, and then we quench $V'_{ini}\rightarrow V'$.
After a systematic analysis, we have found that the short-time dynamics depends strongly on the 
initial state and the final Hamiltonian, so we will focus here on the long-time dynamics and 
the outcome of relaxation. As discussed previously \cite{manmana07,rigol09STATa,rigol08STATc}, 
after relaxation, observables are well described by the diagonal ensemble 
$O_{diag}=\sum_{\alpha} |C_{\alpha}|^{2} O_{\alpha\alpha}$, where $C_{\alpha}$ is the overlap of 
the initial state with eigenstate $\alpha$ of the Hamiltonian.

\begin{figure}[!h]
\begin{center}
\includegraphics[width=0.47\textwidth,angle=0]{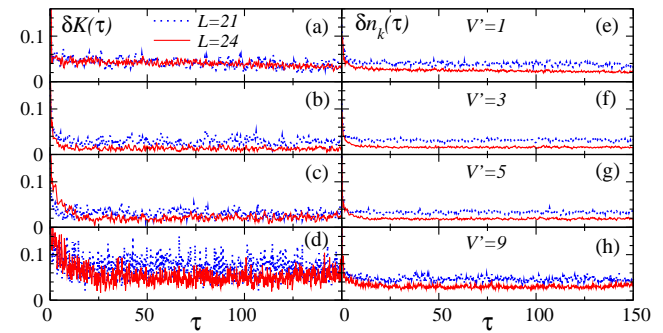}
\end{center}
\vspace{-0.6cm}
\caption{\label{Fig:TimeEvolution}(Color online)
Dynamics of the normalized difference between the evolving expectation values of $K$ (left panels)
and $n(k)$ (right panels) and the diagonal ensemble prediction. Average over the evolution of nine 
initial states selected from the eigenstates of the Hamiltonian with $V'_{ini}=0,1,\ldots,9$ 
(excluding the $V'$ used for the dynamics). The nine states for each $V'$ were chosen such that 
the (conserved) energies during the time evolution are the same in all cases, and $T=3$. 
Given the energy of the initial state in the final Hamiltonian 
$E=\langle\psi_{ini}\vert \widehat{H}\vert \psi_{ini}\rangle$, 
$T$ is computed by following Ref.\ \cite{temperature}.}
\end{figure}

In Fig.\ \ref{Fig:TimeEvolution}, we show the normalized difference between the time-evolving 
expectation value of $K$ and $n(k)$ and the diagonal ensemble prediction 
$\delta K$ (left panels) and $\delta n_k$ (right panels), respectively. We define
$\delta K(\tau)=|K(\tau)-K_{diag}|/|K_{diag}|$ and 
$\delta n_k(\tau)=(\sum_k|n(k,\tau)-n_{diag}(k)|)/(\sum_k n_{diag}(k))$.
In order to verify the universality of our results, for each value of $V'$ used for the 
dynamics, we prepared nine initial states selected from the eigenstates of the Hamiltonian 
with different values of $V'_{ini}$ (excluding $V'$) and studied the dynamics for all of 
them. The long time dynamics was found to be very similar, independent of the initial state. 
In Fig.\ \ref{Fig:TimeEvolution}, we depict the average over those nine different 
time evolutions \cite{supplement}. These plots show that after long times, observables relax 
to values similar to those predicted by the diagonal ensemble, and those predictions become 
more accurate with increasing system size. Only for $V'=9$ do we find large time fluctuations 
of $K$, which is a consequence of the approach to localization in $k$ space. 
However, even these time fluctuations decrease with increasing system size.

Once it is known that the relaxation dynamics brings observables to the values predicted 
by the diagonal ensemble, and that the accuracy of that prediction improves with increasing 
system size, all we need to do to check whether the system thermalizes or not is to compare 
the predictions of the diagonal ensemble with the microcanonical ones. For larger systems, 
one could also compare with the canonical ensemble, but this is 
not adequate here due to finite-size effects \cite{rigol09STATa}. 

\begin{figure}[!h]
\begin{center}
\includegraphics[width=0.47\textwidth,angle=0]{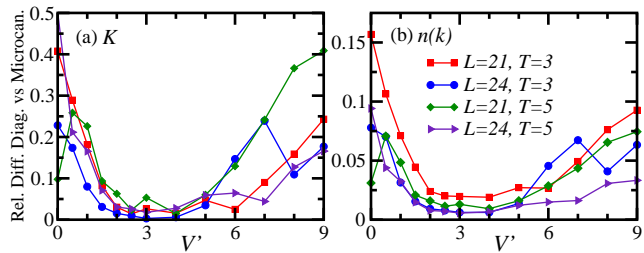}
\end{center}
\vspace{-0.6cm}
\caption{\label{Fig:Thermodynamics}(Color online)
Relative differences between the predictions of the microcanonical and diagonal ensembles for 
(a) $K$ and (b) $n(k)$ vs $V'$, for $T=3.0$ and $T=5.0$. Results are shown for $L=21$, $N_b=7$ 
and  $L=24$, $N_b=8$. The diagonal ensembles correspond to the quenches from the nine 
$V'_{ini}$ of Fig.\ \ref{Fig:TimeEvolution} to $V'$, and the curves are averages over
the nine relative differences. Relative differences are computed in exactly the 
same way as $\delta K$ and $\delta n_k$ in Fig.\ \ref{Fig:TimeEvolution}.
}
\end{figure}

In Fig.\ \ref{Fig:Thermodynamics}, we depict the comparison between diagonal and 
microcanonical ensembles. Results are shown for the same set of quenches 
and initial states presented in Fig.\ \ref{Fig:TimeEvolution}, and for an additional set 
of initial states such that the effective temperature of the relaxed systems is a bit higher, 
namely, $T=5$ \cite{supplement}. The results for $T=3$ and $T=5$ are in qualitative agreement 
with each other for our two observables of interest. They show that the microcanonical ensemble 
predicts the outcome of the relaxation dynamics with high accuracy for the intermediate values 
of $V'$, where ETH was shown to be valid (cf. Fig.\ \ref{fig:eth}) and where quantum chaos was 
seen to emerge (cf. Fig.\ \ref{fig:chaos}). It also shows that (i) the predictions of the 
microcanonical ensemble become more accurate with increasing system size, and (ii) the 
increasing disagreement between the microcanonical prediction and the outcome of the 
relaxation dynamics, after crossing the superfluid to insulator transition, moves to larger 
values of $V'$ as the system size increases.

To summarize, our studies indicate that thermalization does occur in gapped systems.
If integrability is broken, ETH was shown to be valid, away from the edges of the spectrum, 
even if gaps are present in the spectrum and the ground state of the system is an insulator. 
We verified that ETH holds where quantum chaos develops and that thermalization closely follows 
the validity of ETH (i.e., we found no instance where the rare event scenario put forward in 
Ref.\ \cite{biroli09} emerges in these systems). Our analysis of different lattice sizes showed 
that: (i) The range of parameters over which ETH applies increases with increasing system size; 
in particular, ETH becomes valid deeper into the insulating side of the phase diagram, and 
(ii) away from integrability, the fluctuations of the eigenstate expectation values 
of few-body observables decrease with increasing systems size. Further studies are needed to 
understand the dependence of the short-time dynamics on the initial state as well as the final 
Hamiltonian, and to determine the precise scaling, with system size, for the 
onset of ETH and quantum chaos.

\vspace{-0.35cm}

\begin{acknowledgments}
This work was supported by ONR (M.R.) and by the Research Corporation (L.F.S.). We thank 
Amy Cassidy and Maxim Olshanii for useful comments about the manuscript.
\end{acknowledgments}

\onecolumngrid

\vspace*{0.4cm}

\begin{center}

{\large \bf Supplementary material for EPAPS
\\ Quantum chaos and thermalization in gapped systems}\\

\vspace{0.6cm}

Marcos Rigol$^1$ and Lea F. Santos$^2$\\

$^1${\it Department of Physics, Georgetown University, Washington, DC 20057, USA}

$^2${\it Department of Physics, Yeshiva University, New York, NY 10016, USA}
 
\end{center}

\vspace{0.6cm}

\twocolumngrid

\section{Hamiltonian and gapless-gapped transition}

We study the one-dimensional hard-core boson (HCB) model with nearest-neighbor (NN) 
hopping $t$, and nearest- and next-nearest-neighbor (NNN) interaction $V$ and $V'$, 
respectively.  The Hamiltonian is written as
{\setlength\arraycolsep{0.5pt}
\begin{eqnarray}
&&\hat{H}_{b}=\sum_{i=1}^L \left\lbrace -t\left( \hat{b}^\dagger_i \hat{b}_{i+1} 
+ \textrm{H.c.} \right) \right. \label{Eq:hamiltonianHCB_suppl} \\
&&+V\left.\left( \hat{n}^b_i-\dfrac{1}{2}\right)\left( \hat{n}^b_{i+1}-\dfrac{1}{2}\right) 
 +V'\left( \hat{n}^b_i-\dfrac{1}{2}\right)\left( \hat{n}^b_{i+2}-\dfrac{1}{2}\right)\right\rbrace. 
\nonumber
\end{eqnarray}
}Above, we take $\hbar =1$, $L$ is the size of the chain, $\hat{b}_i$ ($\hat{b}_i^{\dagger}$ ) is the 
bosonic annihilation (creation) operator on site $i$ and $\hat{n}_i^b= \hat{b}_i^{\dagger} \hat{b}_i$ 
is the boson local density operator. HCBs do not occupy the same site, so $b_i^2=b_i^{\dagger 2}=0$.

Hamiltonian~(\ref{Eq:hamiltonianHCB_suppl}) conserves the total number of particles $N_b$ and 
is translational invariant; it is then composed of independent blocks each associated with 
a value of $N_b$ and a total momentum $k$. Here we select $N_{b}=L/3$ and study all values 
of $k$, from 0 to $\ldots \left\lfloor L/2 \right\rfloor$. Lattices with up to 24 sites and 
8 HCBs are considered, corresponding to a total Hilbert space of dimension $D=735\,471$. 
The dimension $D_k$ of each $k$-sector is given in Table \ref{table:dimensions}. 
We perform a full exact diagonalization of each sector independently.

\begin{table}[h]
\caption{Dimension of subspaces}
\begin{center}
\begin{tabular}{|c|c|c|c|c|}
\hline
\hline 
$L=18$ & $k=0,6 $ &$k=1,5,7 $ & $k=2,4,8 $  & $k=3,9$  \\
$D_k$ & 1038 & 1026 & 1035 & 1028 \\
\hline
$L=21$ & $k=0,7 $ & all other $k$'s & &  \\
$D_k$  & 5538 & 5537 &  & \\
\hline
$L=24$ & $k=0,8 $ & odd $k$'s & $k=2,6,10 $ & $k=4 $  \\
$D_k$ & 30667 & 30624 & 30664 & 30666 \\
\hline 
\hline
\end{tabular}
\end{center}
\label{table:dimensions}
\end{table}

In what follows, $t=1$ sets the energy scale and we only consider repulsive interactions 
$V, V' >0$. We fix $V=6$ and vary $V'$ from 0 to 9. The system is integrable when $V'=0$, 
while the addition of NNN interaction may lead to the onset of chaos. In addition, 
there is a critical value of the NNN interaction, $V'_c=3$, below which  the ground state 
is a gapless superfluid and above which it becomes a gapped insulator 
(see Ref.\ [17] in main text).

\section{Quantum Chaos Indicators}

To analyze the transition from integrability to chaos, we study chaos indicators
that depend only on the eigenvalues of the system, such as level spacing distribution and 
level number variance, and indicators that measure the level of delocalization of the eigenvectors, 
such as the inverse participation ratio (IPR) and the information (Shannon) entropy (S).

\subsection{Level spacing distribution and level number variance}

Level spacing distribution and level number variance give information about 
short-range and long-range correlations, respectively. They are obtained from the 
unfolded spectrum of each symmetry sector separately. The procedure of unfolding consists 
of locally rescaling the energies, so that the mean level density of the new sequence of 
energies is 1.

For integrable systems, the distribution of spacings $s$ of neighboring energy levels 
may cross and the distribution is Poissonian, 
\[
P_{P}(s) = \exp(-s),
\] 
whereas in nonintegrable systems, level repulsion leads to the Wigner-Dyson distribution. 
The form of the latter depends on the symmetries of the system. 
Hamiltonian~(\ref{Eq:hamiltonianHCB_suppl}) is time-reversal and rotationally invariant, 
it therefore gives the same distribution as random matrices with real and symmetric elements, 
the so-called Gaussian Orthogonal Ensembles (GOEs):
\[
P_{\text{GOE}}(s) = \frac{\pi s}{2}\exp\left(-\frac{\pi s^2}{4}\right).
\]

The level number variance is defined as
\[
\Sigma^2(l) \equiv \langle (N(l,\epsilon)^2 \rangle - \langle N(l,\epsilon)\rangle^2,
\]
where $N(l,\epsilon)$ is the number of states in the energy interval $[\epsilon,\epsilon+l]$
and $\langle . \rangle $ is the average over different initial values of the energy level 
$\epsilon$. For a Poisson distribution, 
\[
\Sigma^2_{P}(l)=l,
\] 
while for GOEs in the limit of large $l$,
\[
\Sigma^2_{\text{GOE}}(l)=\frac{2}{\pi^2}\left[\ln (2\pi l) + \gamma +1 -\frac{\pi^2}{8} \right],
\] 
where $\gamma$ is the Euler constant.

\begin{figure}[htb]
\includegraphics[width=0.47\textwidth]{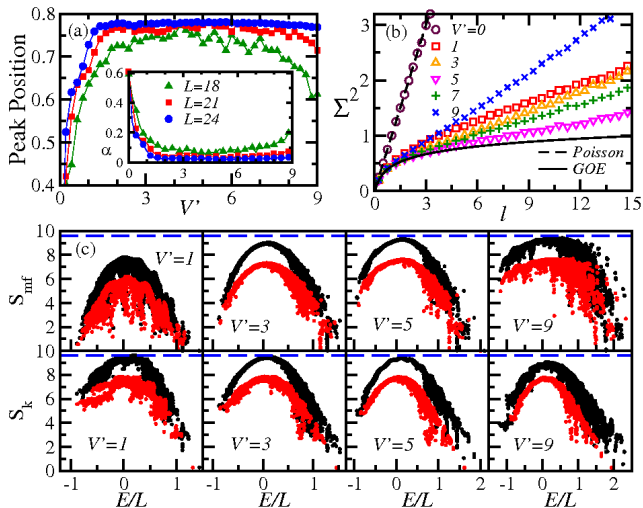}
\vspace{-0.25cm}
\caption{(Color online)
(a) Peak position of the level spacing distribution averaged over all sectors with 
$k=1, \ldots, \left\lfloor L/2 \right\rfloor$. The inset shows the quantity 
$\alpha$ (see text). (b) Level number variance averaged over the same $k$-sectors, 
for $L=21$, and compared to the Poisson and GOE results. (c) Shannon entropy in the 
mean-field (top) and momentum (bottom) basis vs energy per site; $k=2$; black: $L=24$, 
red: $L=21$;  dashed line: S$_{\text{GOE}} \approx 9.5969$ for $D_2=30\,664$.}
\label{fig:chaos_supp}
\end{figure}

Results for level spacing distribution and level number variance for various values of 
$V'$ are shown in Figs.~\ref{fig:chaos_supp}(a) and \ref{fig:chaos_supp}(b). These figures 
are equivalent to Figs.~1(a) and 1(b) in the main text. Note that because parity is a 
symmetry found only in the sector with $k=0$, this subspace is not included in the averages 
for both figures.

In Fig.~1(a) of the main text we quantified the integrable-chaos transition in terms
of the parameter $\beta$ of the Brody distribution. In Fig.~\ref{fig:chaos_supp}(a) 
we consider the peak position of the distribution, which shifts from zero to $\approx 0.7979$ 
during the crossover from $P_P(s)$ to $P_{WD}(s)$, and a quantity $\alpha$ (inset) defined as
\begin{equation}
\alpha \equiv \frac{\sum_i |P(s_i)-P_{WD}(s_i)|}{\sum_i P_{WD}(s_i)}
\label{alpha}.
\end{equation}
Above the sum runs over the whole spectrum. In the chaotic limit $\alpha \rightarrow 0$.

Figures~\ref{fig:chaos_supp}(a) and \ref{fig:chaos_supp}(b) reinforce the main results of 
Figs.~1(a) and 1(b) in the main text, which are: (i) Two transitions are seen, from integrability 
to chaos as $V'$ increases from $V'=0$, and a departure from chaoticity for large values of $V'$.
(ii) By comparing the values of $V'$ that induce chaos with the $V'$ for the superfluid-insulator 
transition in the inset of Fig.~1(a), a region of overlap between the gapped phase and the 
chaotic regime is identified. (iii) A direct dependence exists between the system size and
the width of the interval of $V'$ values that lead to chaos; for larger systems the chaotic 
behavior goes deeper into the insulating phase. Larger system sizes bring also better agreement 
with the GOE results, as seen by comparing  Fig.~1(b) in the main text, which was obtained for
24 sites, with Fig.~\ref{fig:chaos_supp}(b), which deals with $L=21$.

\subsection{Delocalization Measures}

Delocalization measures, such as IPR and S, quantify the level of complexity of the eigenvectors 
(see references in main text). In general, they depend on the basis in which the computations 
are performed. For an eigenstate $|\psi_{\alpha}\rangle$ of Hamiltonian (\ref{Eq:hamiltonianHCB_suppl}) 
written in the basis vectors $|\phi_j\rangle$ as $|\psi_{\alpha}\rangle = \sum_{j=1}^{D_{k}} c^j_{\alpha} 
|\phi_j\rangle$, IPR and S are respectively given by
\[
\mbox{IPR}_{\alpha} \equiv \left(\sum_{j=1}^{D_k} |c^j_{\alpha}|^4\right)^{-1}
\]
and
\[
\mbox{S}_{\alpha} \equiv -\sum_{j=1}^{D_k}  |c^j_{\alpha}|^2 \ln |c^j_{\alpha}|^2.
\]
These quantities indicate how much spread each $|\psi_{\alpha}\rangle$ is in the selected basis.

In the case of GOEs, the eigenvectors do not depend on the basis. They are simply random vectors,
which give $\mbox{IPR}_{\text{GOE}} \sim D_k/3$ and $\mbox{S}_{\text{GOE}} \sim \ln(0.48 D_k)$. 
An essential difference of our system with respect to GOEs, besides the absence of randomness on it, 
is that Hamiltonian (\ref{Eq:hamiltonianHCB_suppl}) has only two-body interactions. Consequently, 
its eigenstates may approach the GOE results only in the middle of the spectrum, close to the edges 
chaos does not fully develop.

In Fig.~\ref{fig:chaos_supp}(c), we show S for two system sizes and in two bases. In the 
mean-field (mf) basis, $|\phi_j\rangle$'s correspond to the eigenstates of the integrable 
Hamiltonian ($V'=0$); this choice separates regular from chaotic behavior. In the $k$-basis, 
$|\phi_j\rangle$'s are the basis vectors of total momentum. As $V'$ increases from zero and the 
system undergoes the first transition, from integrability to chaos, the values of S$_{\text{mf}}$
become larger, indicating delocalization  of the eigenvectors in the mf-basis. The second transition, 
signaling the departure from chaoticity, is followed by the reduction of the values of S$_{k}$ and 
the consequent localization of the eigenvectors in the $k$-basis. Even though the shrinking of the 
eigenvectors in $k$-space is better visualized in terms of IPR (main text), it is important to 
observe again that, similarly to the distancing of the eigenvalues from $P_{WD}(s)$ and 
$\Sigma^2_{\text{GOE}}(l)$, this localization occurs when $V'$ is already beyond
the critical point $V'_c$ for the gapless-gapped transition.

The structure of the eigenstates gives us information about what to expect in terms of 
thermalization. The eigenstate thermalization hypothesis (ETH) states that thermalization 
should happen when the eigenstate expectation values (EEVs) do not fluctuate for states 
which are close in energy. In this case, the EEVs will equal the microcanonical average. 
The validity of ETH is therefore certain to hold for GOEs, where all eigenstates are extended 
and have the same level of complexity. For our two-body interaction system, two aspects must 
be taken into account: chaoticity and the energy of the initial state. Away from the chaotic 
regime, that is when $V' \rightarrow 0$ or $V' \gg V'_c$, the structure of eigenstates close 
in energy fluctuate significantly, as shown by the panels of Fig.~\ref{fig:chaos_supp}(c).
But even in the chaotic region, fluctuations are seen also at the edges of the spectrum.
Thus, thermalization in our lattice should occur in the chaotic domain and for initial states 
with energy away from the spectrum borders. This conclusion is the same we arrived at in 
Ref.\ [16] (main text), where we studied systems whose ground state was always gapless;
therefore, as expected, the superfluid-insulator transition (a quantum phase transition)  
does not appear to affect the level of complexity of the bulk states. We note, however, 
that the results of the delocalization measures for system (\ref{Eq:hamiltonianHCB_suppl}) 
show an asymmetry with respect to the middle of the spectrum that was not so evident in 
the systems of Ref.\ [16] (cf. Fig.~1(c), Fig.~\ref{fig:chaos_supp}(c), and Figs.11-16 
from Ref.\ [16]). In the present lattice, larger fluctuations in the values of S and IPR are 
present for $E>0$ for all values of $V'$ studied.

\section{Eigenstate Thermalization Hypothesis}

The connection between the ETH and the chaos indicators is reinforced by comparing the 
outcomes for the delocalization measures with those for the EEVs.

\begin{figure}[htb]
\includegraphics[width=0.47\textwidth]{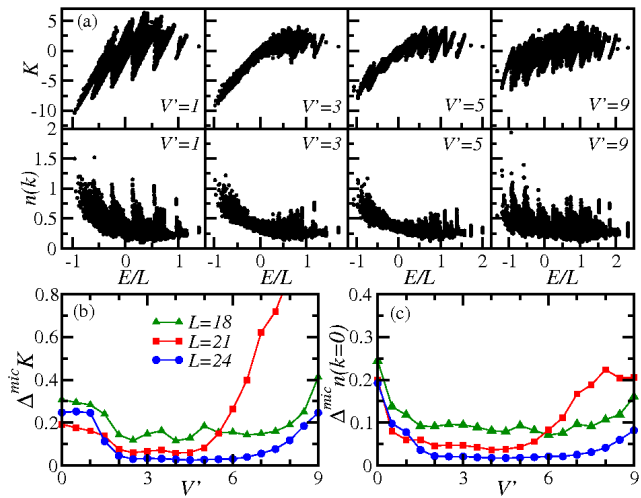}
\vspace{-0.25cm}
\caption{(Color online)
(a) Eigenstate expectation values (EEVs) of $K$ (top) and $n(k)$ (bottom) vs energy per site for 
the full spectrum (including all momentum sectors). Results are shown for four different 
values of $V'$ and $L=21$. Panels (b) and (c) correspond to the average relative deviation 
of the EEVs of $K$ (b) and $n(k=0)$ (c) with respect to the microcanonical result vs $V'$ 
for $T=5$ (see text) and three different lattice sizes. $T=5$ ($L=24$) corresponds to 
$E=-6.24$ for $V'=1$, $E=-5.55$ for $V'=3$, $E=-6.61$ for $V'=5$, and $E=-12.68$ for $V'=9$.}
\label{fig:eth_supp}
\end{figure}

The top and bottom panels of Fig.~\ref{fig:eth_supp}(a) show, respectively, the 
EEVS of the kinetic energy 
\[
\hat{K}=\sum_i -t\left( \hat{b}^\dagger_i \hat{b}_{i+1} + \textrm{H.c.} \right)
\] 
and the momentum distribution function 
\[
\hat{n}(k)=\frac{1}{L} \sum_{i,j} e^{-ik(i-j)} \hat{b}_i^{\dagger} \hat{b}_j
\] 
for all eigenstates. Fig.~\ref{fig:eth_supp}(a) is analogous to Fig.~2(a), but refers 
to a smaller system. The results in both figures mirror the behavior of the delocalization measures 
throughout the transitions. Large fluctuations of the EEVs are seen over the entire spectrum 
when the system is away from the chaotic limit, that is, when it approaches integrability and 
when it approaches localization in $k$-space. When chaos sets in, the fluctuations decrease 
mainly in the center of the spectrum, where the ETH is expected to be valid. We also should stress that, 
by comparing Fig.~\ref{fig:eth_supp}(a) with Fig.~2(a) in the chaotic regime, one can see that the 
fluctuations of the observables in {\em all} individual eigenstates (away from the edges of the spectrum) 
decrease with increasing systems size. This is a clear indication that the rare state scenario 
introduced in Ref.\ [13] in the main text does not take place in these systems.

To quantify the deviation of the EEV for an observable $O$ with respect to the microcanonical 
result ($\Delta^{mic}$), we define
\[
\Delta^{mic}O \equiv \frac{\sum_{\alpha}\,
|O_{\alpha \alpha}-O_{mic}|}{\sum_{\alpha}\,O_{\alpha \alpha}}.
\] 
Above, the sum runs over the microcanonical window, $O_{\alpha\alpha}$ are the EEVs of 
the operator $\hat{O}$, and the microcanonical expectation values $O_{mic}$ are computed as usual. 

Figures\ \ref{fig:eth_supp}(b) and \ref{fig:eth_supp}(c) show the relative deviation
$\Delta^{mic}K$ and $\Delta^{mic}n(k=0)$  averaged over all momentum sectors and for 
all eigenstates that lie within a window $[E-\Delta E, E+\Delta E]$, where $\Delta E=0.1$. 
We select $E$ according to the effective temperature $T$ that we want to study [$T=5$ here
and $T=3$ in the equivalent Figs.~2(b) and 2(c) in the main text]. The analysis in terms of 
a single temperature allows for a fair comparison of all systems sizes and values of $V'$. 
Temperature and energy are related by the expression
\begin{equation}
E=\dfrac{1}{Z}\textrm{Tr}\left\lbrace \hat{H} e^{-\hat{H}/{k_B T}}\right\rbrace,
\label{EtoT}
\end{equation}
where $Z=\textrm{Tr}\left\lbrace e^{-\hat{H}/{k_BT}}\right\rbrace$ is the partition function, 
and we set $k_B$ to unity.

The results for $\Delta^{mic}K$ and $\Delta^{mic}n(k=0)$ for any given systems size comply with the predictions of 
chaos measures. As the system size increases, the average deviations for both observables decrease 
and the width of the interval of values of $V'$ for which the EEVs approach the thermal average 
increases. We should add that for some specific values of $V'$, the relative deviations are seen to be 
larger in some larger system sizes, that is, the curves in Fig.\ \ref{fig:eth_supp} 
and Fig. 2 in the main text cross. As we show next, this is a finite size effect related 
to bands of eigenvalues that move, within the temperature scale considered in this work, 
as the system size is changed.

\section{Fluctuation of the EEV's as a function of temperature}

The study of the fluctuation of the EEV's as a function of temperature
gives further support to the discussions above.

\begin{figure}[!h]
\begin{center}
\includegraphics[width=0.47\textwidth,angle=0]{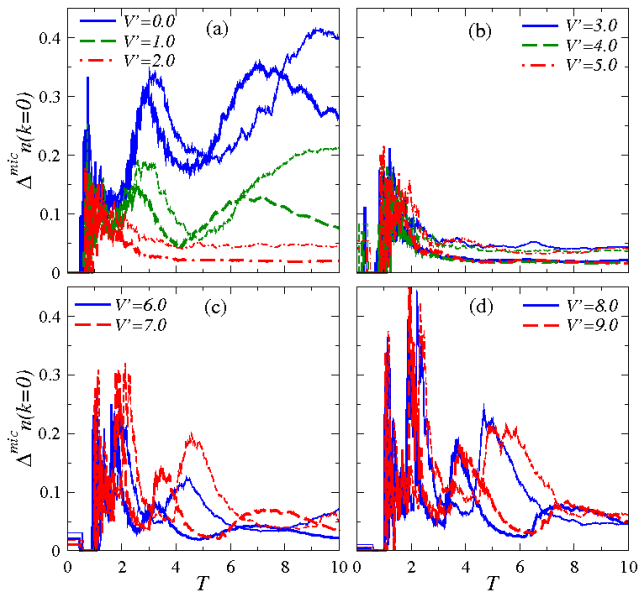}
\end{center}
\vspace{-0.6cm}
\caption{\label{Fig:Deviations}(Color online)
Average relative deviation of eigenstate expectation values
with respect to the microcanonical prediction as a function of the
effective temperature $T$ of the eigenstates.
Results for $\Delta^{mic}n(k=0)$ are given for the
nine values of $V'$ indicated and for systems with $L=24$ (thick lines) and 
$L=21$ (thin lines).}
\end{figure}

In Fig.~\ref{Fig:Deviations}, we present results for $\Delta^{mic}n(k=0)$ vs $T$
for nine different values of $V'$ for systems with 24 and 21 lattice sites 
and $T\leq 10$. As before, distinct features are associated with different regimes.
Far from chaoticity, large values of $\Delta^{mic}n(k=0)$ appear for all temperatures 
considered. In the chaotic domain, on the other hand, large values of $\Delta^{mic}n(k=0)$ 
are restricted to low temperatures, while at large $T$, $\Delta^{mic}n(k=0)$ saturates 
at small values. This corroborates our statements that the validity of ETH for these finite
systems goes hand in hand with the onset of chaos and holds away from the edges of the spectrum. 

In general, we also observe that larger systems reduce the average relative deviations.
However, especially away from chaoticity, this rule may not be obeyed for particular values 
of the effective temperature and particular values of $V'$. Notice in Fig.~\ref{Fig:Deviations} 
the peaks with large fluctuations that move within our temperature scale as the system 
size is increased. These are responsible for the bumps seen in Figs.\ 2(b) and 2(c) in the 
main text and for the behavior of the fluctuations of the system with $L=21$ in 
Figs.\ \ref{fig:eth_supp}(b) and \ref{fig:eth_supp}(c). From the behavior seen here with 
increasing systems size, we expect them to disappear for very large system sizes.

\section{Long-Time Dynamics after a Quantum Quench}

The endorsement of the predictions for thermalization drawn from chaos measures and the ETH
is done in two steps. First we study the relaxation dynamics of the system to verify that 
it brings the observables $O$ close to the values of the diagonal ensemble, 
\[
O_{diag}=\sum_{\alpha} |C_{\alpha}|^{2} O_{\alpha\alpha},
\] 
where $C_{\alpha}$ is the overlap of the initial state $|\Psi(0)\rangle$ with the eigenstate 
$\alpha$ of the Hamiltonian, $C_{\alpha}=\langle \psi_{\alpha} |\Psi(0)\rangle$. Then
we compare the results for the diagonal ensemble with those for the microcanonical ensemble. 
When they coincide we say that thermalization has taken place.

In the main text, the second step was presented in Fig.~4 for two values of temperature,
$T=3$ and $T=5$. The relative differences between the predictions of the microcanonical and 
diagonal ensembles for $K$ and $n(k)$ became indeed small for values of $V'$ where the ETH 
was shown to be valid. The first step was illustrated in Fig.~3 for $T=3$ and, for completeness, 
we now demonstrate that similar results hold for the dynamics of systems for 
which the effective temperature is $T=5$.

\begin{figure}[!h]
\begin{center}
\includegraphics[width=0.47\textwidth,angle=0]{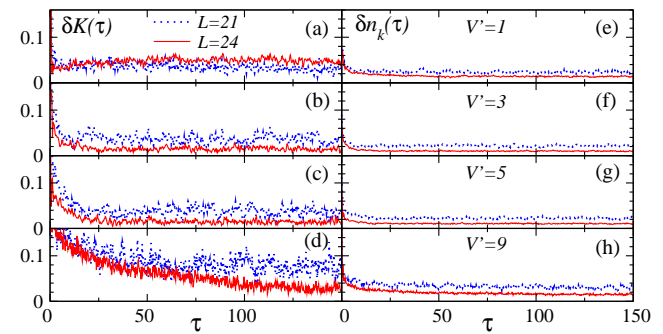}
\end{center}
\vspace{-0.6cm}
\caption{\label{Fig:TimeEvolution_supp}(Color online)
Dynamics of the normalized difference between the evolving 
expectation values of $K$ (left panels)
and $n(k)$ (right panels) and the diagonal ensemble prediction.
The plots are the result of the average over the evolution of nine initial states selected from 
the eigenstates of the Hamiltonian with $V'_{ini}=0,1,\ldots,9$ (excluding the $V'$ used for the 
dynamics). The nine states for each $V'$ were chosen such that the (conserved) energies 
during the time evolution are the same in all cases, and the effective temperature $T=5$. 
Given the energy of the initial state in the final Hamiltonian 
$E=\langle\psi_{ini}\vert \widehat{H}\vert \psi_{ini}\rangle$, the effective temperature 
is computed following Eq.~(\ref{EtoT}).}
\end{figure}

Figure\ \ref{Fig:TimeEvolution_supp} gives the normalized difference between the time 
evolving expectation value of $K$ and $n(k)$ and the diagonal ensemble prediction, 
$\delta K$ (left panels) and $\delta n_k$ (right panels), respectively. We define
$\delta K(\tau)=|K(\tau)-K_{diag}|/|K_{diag}|$ and 
$\delta n_k(\tau)=(\sum_k|n(k,\tau)-n_{diag}(k)|)/(\sum_k n_{diag}(k))$.
For each value of $V'$ considered in the analysis of the dynamics, nine initial states, 
all corresponding to $T=5$, were selected from the eigenstates of the Hamiltonian with 
different values of $V'_{ini}$ (excluding $V'$). We studied the evolution of the nine
states after the quench and verified that their long-time dynamics were very similar.
In Fig.\ \ref{Fig:TimeEvolution_supp}, we display the average over those nine different 
time evolutions. The plots show that after long times, observables relax to values similar to 
those predicted by the diagonal ensemble and those predictions become more accurate with 
increasing system size. Only for $V'=9$ we find large time fluctuations of $K$ and a much 
longer time scale for relaxation, which is a consequence of the approach to localization 
in $k$-space. However, even in the latter case, the time fluctuations are always seen to 
decrease with increasing system size.


\begin{thebibliography}{30}
\bibitem{temperature} Temperature ($T$) and energy ($E$) are related by 
$E=\dfrac{1}{Z}\textrm{Tr}\left\lbrace \hat{H} e^{-\hat{H}/{k_B T}}\right\rbrace$,
where $Z=\textrm{Tr}\left\lbrace e^{-\hat{H}/{k_BT}}\right\rbrace$, $k_B=1$.
\end{thebibliography}
\end{document}